\documentclass[
reprint,
amsmath,amssymb,
aps,
prl,
longbibliography,
floatfix
]{revtex4-2}

\usepackage[english]{babel}

\usepackage{mathrsfs}

\usepackage{graphicx}% Include figure files
\usepackage[T1]{fontenc}
\usepackage[utf8]{inputenc}
\usepackage[table, dvipsnames]{xcolor}		% great colors (dvipsnames for more choice)
\usepackage{verbatim}
\usepackage{dcolumn}% Align table columns on decimal point
\usepackage{hyperref}
\usepackage[capitalise]{cleveref}
\usepackage[mathlines]{lineno}% Enable numbering of text and display math
\usepackage{svg}
\usepackage{siunitx}						% formatted units
\usepackage{mathtools}						% improves formatting of amsmath-stuff
\usepackage{physics} 						% for other things
\usepackage{braket} 						% for \braket-command (handy in QM)
\usepackage{MnSymbol}
\usepackage{enumitem}                       % itemize / enumerate adjustments
\usepackage{multirow}                       % Merging cells vertically
\usepackage{array}     % Better column formatting
\usepackage{bbm} % For mathbbm{1}

\renewcommand{\i}{\mathrm{i} }
\renewcommand{\d}{\mathrm{d} }
\newcommand{\e}{\mathrm{e}}
\newcommand{\Ae}{\mathrm{AMP}}
\newcommand{\QDAE}{{\mathrm{QD}\text{-}\mathrm{AMP}}}
\newcommand{\efield}{\mathcal{E}}

\newcommand{\eff}{\mathrm{eff}}

\newcommand{\rar}{\rightarrow}

\renewcommand{\H}{\mathcal{H}}

\renewcommand{\P}{\mathcal{P}}
\newcommand{\BH}{\mathrm{BH}}
\newcommand{\RS}{\mathrm{RS}}

\newcommand{\Q}{\mathcal{Q}}

\newcommand{\wave}{\hat{\Omega}}
\newcommand{\ddt}{\frac{\mathrm{d}}{\mathrm{d}t}}

\DeclareSIUnit\fs{fs}						% defines femtosecond
\DeclareSIUnit\eV{eV}						% defines electron volt
\DeclareSIUnit\au{a.u.}						% defines electron volt

\newsavebox{\mytablebox}

\begin{document}

\preprint{APS/123-QED}

%\title{Effective Hamiltonians in intense light-matter interaction: Application of quasi-degenerate Rayleigh-Schrödinger perturbation theory}
%\title{Perturbative effective Hamiltonians in intense light-matter interaction}
%\title{Reconciliation of effective Hamiltonians for intense light-matter interaction using Rayleigh-Schrödinger perturbation theory}
\title{Reconciliation of effective Hamiltonians for intense light-matter interaction}

\author{Jakob Nicolai Bruhnke%\orcidlink{0000-0001-7736-5620}
}
\author{Jan Marcus Dahlström$^{\text{\dagger}}$%\orcidlink{0000-0002-5274-1009}
}

\affiliation{Department of Physics, Lund University, 22100 Lund, Sweden. }

\begin{abstract}
\noindent 
Essential-state models are central for quantum control and technology in broad regimes of light-matter interaction. The {canonical} effective Hamiltonian is obtained equivalently from adiabatic elimination, the Markov approximation, and the pole approximation. These approximations are known to break down at high intensities, significantly limiting their applicability {to moderate} light-matter interaction. We show how this limitation can be addressed by applying quasi-degenerate Rayleigh-Schrödinger perturbation theory (QD-RSPT). We reconcile QD-RSPT with adiabatic elimination and propose a quasi-degenerate extension of adiabatic elimination that is robust when the detuning of the essential states is non-negligible. The accuracy of QD-RSPT is demonstrated in both the low- and high-frequency regime, showing excellent agreement with Floquet calculations {at high intensities}. The crucial corrections to adiabatic elimination make the eigenvectors of the effective Hamiltonian non-orthogonal. Physically, this allows us to account for the asymmetric strength with which different essential states couple to the non-essential states. We expect {that our} systematic approach to effective Hamiltonians from QD-RSPT {will} constitute a new state of the art in intense light-matter interaction {and quantum optics with novel forms of strong coupling and quantum control phenomena}  being conceivable.
\end{abstract}
  
\maketitle

\begin{table}[b!]
\begin{flushleft}
  $^\text{\dagger}$ marcus.dahlstrom@fysik.lu.se
\end{flushleft}
\end{table}

\twocolumngrid

\section{I. Introduction}

Coherent control is centered around the goal of precisely controlling electrons in quantum systems such as atoms, molecules, and condensed matter. In the low-frequency regime, where the photon energy is small compared to the binding energy, this control is routinely exerted through coherent driving of bound-bound and bound-continuum transitions by intense fields, giving rise to a rich landscape of non-linear quantum phenomena. These include the Autler-Townes-like splittings in resonant multiphoton ionization of alkali atoms \cite{wollenhauptControlInterferencesAutlerTownes2003} and the associated spin-orbit dynamics \cite{bayerTimeresolved3DImaging2019, liPartialWaveResolvedSpinOrbit2025}, with recent work exploring spin polarization \cite{zhangSpinPolarizationStrongField2025}, signatures beyond the rotating-wave approximation (RWA) \cite{anandAttosecondCounterrotatingwaveEffect2017}, coherent control of Stark shifts through dynamic interference \cite{bertolinoThomasReicheKuhnCorrectionTruncated2022, vismarraetal.DynamicInterferenceChirped2025, liObservationLaserAssistedDynamic2024}, and two-photon Rabi oscillations in alkali atoms \cite{tothProbingStrongfieldTwophoton2021}. %, and dynamic interference  \textbf{vector light, time delays, etc. (OPEN TO SUGGESTIONS)}. %Since (1+2)-REMPI from alkali ground states leads through the Rydberg states, complex quasi-energy structures emerge, where ionization can, e.g., be suppressed through anti-resonances \textbf{cite ourselves in the future}. Alternatively, (1+2)-REMPI processes in 
As the laser field strength is increased further, we enter the strong-field regime of light-matter interaction, in which we stop thinking about electronic populations in field-free states and start thinking in terms of electronic trajectories, leading to high-harmonic generation (HHG) \cite{ferrayMultipleharmonicConversion10641988} and attosecond physics \cite{krauszAttosecondPhysics2009}.

%In the lower frequency regimes, quantum control has long been established. Quantum logic gates are commonly implemented through a two-level system in a resonant electromagnetic field \textbf{CITE SOMEONE}. Other quantum computing platforms exploit Raman transitions for precise population transfer \textbf{CITE ANDREAS WALTER}. Two-photon transitions are routinely leveraged to populate high Rydberg states \cite{kumlinQuantumOpticsRydberg2023}. In strong-field physics, Rabi oscillations in Alkali atoms have been driven with infrared (IR) fields. 

%There, the Autler-Townes-like doublet due to resonance-enhanced multiphoton ionization (REMPI) has been studied \textbf{CITE WOLLENHAUPT}, and more recently effects due to spin-orbit splitting \textbf{CITE WOLLENHAUPT}. In quantum computing, increasing fault tolerance is a critical problem. In order to accurately describe the population dynamics, the corrections beyond the rotating-wave approximation (RWA) have to be accounted for systematically, and the impact of these corrections onto the time-evolution be assessed. Meanwhile, in (1+2) resonance-enhanced multiphoton ionization (REMPI) of alkali atoms, the ionization pathway proceeds through the Rydberg states, which at high intensities can lead to suppression of ionization through anti-resonances \textbf{CITE ZHANG RUBIDIUM PAPER}.

Recent advancements in free electron laser science \cite{emmaetal.FirstLasingOperation2010, allariaetal.HighlyCoherentStable2012, mirianetal.GenerationMeasurementIntense2021} have made it possible to drive coherent processes with extreme ultraviolet (XUV) light  \cite{nandietal.ObservationRabiDynamics2022, nandietal.GenerationEntanglementUsing2024, richteretal.StrongfieldQuantumControl2024}. By now, intensities around $\SI{e14}{\watt\per\square\cm}$ are routinely reached in experiments. At these intensities, non-linear processes can become not only relevant, but even dominant. For example, for the resonant $1s^2 \leftrightarrow 1s4p$ transition, two-photon non-resonant ionization from $1s^2$ is more likely than single-photon ionization from $1s4p$ beyond $\SI{2e13}{\watt\per\square\cm}$ \cite{nandietal.ObservationRabiDynamics2022}. Closely related is the dressed-atom stabilization in helium, in which through interference of the two possible ionization pathways (resonant and non-resonant), one of the dressed states becomes stabilized against ionization \cite{olofssonPhotoelectronSignatureDressedatom2023, olofssonControlPhotoionizationResonant2025}. The most prototypical non-linear effect is surely a two-photon transition. This too can be driven with intense high-frequency fields \cite{dorrTimeEvolutionTwophoton1997, yatsenkoSourceMetastableH2s1999, bruhnkeGiantCounterrotatingOscillations2025, kumarBroadbandFemtosecondLasers2026}. %While population inversion through two-photon Rabi oscillations is generally impossible in the 1s-2s hydrogen system \cite{dorrTimeEvolutionTwophoton1997} (only possible through adiabatic passage techniques \cite{yatsenkoSourceMetastableH2s1999}), we have recently shown that it is feasible in helium from 1s$^2$ to 1s3d \cite{bruhnkeGiantCounterrotatingOscillations2025}. 
Notably, at the intensities required to drive these two-photon transitions at ultrafast time scales ($\leq \SI{100}{\fs}$), the excited electron begins to follow the instantaneous electric field in what is known as counter-rotating oscillations \cite{bruhnkeGiantCounterrotatingOscillations2025}, signifying the breakdown of the rotating-wave approximation (RWA) and the onset of strong-field effects. We also want to mention the work from the Stanford Linear Accelerator Center (SLAC) on X-ray Autler-Townes doublets and Mollow triplets \cite{linkeretal.AttosecondInnershellLasing2025}.

This wealth of non-trivial effects demands sophisticated theoretical machinery. For some problems, one must resort to numerical solutions of the semiclassical time-dependent Schrödinger equation (TDSE), while for others, the well-established essential-states approach has the advantage that the atom is reduced to a few-level system. The latter enables us to gain a quantitative understanding of the dynamics driven by the semiclassical field and connect better to Floquet or quantum optics descriptions \cite{faisalTheoryMultiphotonProcesses1987, shoreTheoryCoherentAtomic1990, cohen-tannoudjiAtomPhotonInteractionsBasic1998}. The effects of other states can sometimes be completely ignored; other times it can be parametrized. The \textit{effective Hamiltonian} $H_\eff$ for this few-level system should then contain all the important physics, and its diagonalization gives rise to combined eigenstates of light and matter, referred to as the dressed states. Thanks to their flexibility, their predictive power, and the ease at which physical interpretations may be assigned to calculations, essential-states approaches have enjoyed popularity for several decades across various research fields. These include the area of quantum control \cite{huangTheoryMultiphotonProcesses2025}, where Raman transitions are exploited for precise population transfer \cite{vitanovPopulationTransferDecaying1997, kinosDesigningGateOperations2021}, and two-photon transitions are leveraged to populate high Rydberg states \cite{kumlinQuantumOpticsRydberg2023}.

Parametrizing the influence of nonessential states routinely proceeds via an approximation known as \textit{adiabatic elimination} \cite{shoreTheoryCoherentAtomic1990}, which has been shown to be equivalent to the \textit{Markov approximation} \cite{paulischAdiabaticEliminationHierarchy2014} and the \textit{pole approximation} \cite{brionAdiabaticEliminationLambda2007}. We will thus refer to them as the AMP approximation, and denote this canonical effective Hamiltonian by $H_\Ae$. %The physics behind the approximation can be seen from at least these three angles: One is to view the nonessential states as a set of stiff springs driven by a slowly-varying driving force. Then, adiabatic elimination implies that only the steady-state DC response matters \cite{allenBroadeningSaturationNphoton1982}. The second angle starts at the formal time-evolution of the nonessential states, which depends on the entire history of the essential states. In the Markov approximation, the nonessential states amplitudes at time $t$ only require knowledge of the essential states amplitudes at time $t$ \cite{baroneExtensionGeometricalRepresentation1973, friedmannEffectiveTwolevelHamiltonian1978, allenBroadeningSaturationNphoton1982}. Lastly, the pole approximation prescribes that the dressing of the nonessential states is unimportant, so that the poles of nonessential states in the resolvent may be neglected \cite{mccleanTheoryResonantTwophoton1978, cohen-tannoudjiAtomPhotonInteractionsBasic1998}.
While in the 1980s, experimental constraints informed the opinion that higher-order corrections to adiabatic elimination were too small to be of interest \cite{bakerNonHermitianQuantumTheory1984}, this position must be reconsidered, with several proposals on higher-order effective Hamiltonians having been put forward in the last decade \cite{torosovAdiabaticEliminationNearly2012, paulischAdiabaticEliminationHierarchy2014, sanzAdiabaticEliminationEffective2016, zlatanovMorrisShoreTransformationNondegenerate2020}. 

However, several questions remain concerning the AMP approximation. One such question is how to choose the ``correct'' interaction picture for the adiabatic elimination or Markov approximation, or, equivalently, how to choose the ``correct'' energy in the pole approximation. In previous studies, guidelines were suggested \cite{brionAdiabaticEliminationLambda2007, paulischAdiabaticEliminationHierarchy2014}, but it is unsatisfactory that such a fundamental question should not have a response based on the theory of effective Hamiltonians. A consequence of projecting the full light-matter Hamiltonian onto a Hilbert subspace is that the resulting effective Hamiltonian has \textit{non-orthogonal} eigenstates \cite{paulischAdiabaticEliminationHierarchy2014}. This is commonly ignored, presumably because the terms that give rise to the non-orthogonality are neglected in adiabatic elimination, and appear only in higher-order effective Hamiltonians \cite{paulischAdiabaticEliminationHierarchy2014}. Here, we will detail their physical significance and interpretation. %Finally, it is well-known how to introduce slowly-varying envelopes to effective Hamiltonians obtained through adiabatic elimination \cite{bakerNonHermitianQuantumTheory1984, hansonManifestationsAtomicCore1997, vitanovPopulationTransferDecaying1997, yatsenkoTwophotonExcitationMetastable2005, tothProbingStrongfieldTwophoton2021, zhangEffectNonresonantStates2022}. Similarly, time-dependence of the field (envelopes, chirps) is naturally accounted for in the theory of time-averaged Hamiltonians \cite{jamesEffectiveHamiltonianTheory2007, gamelTimeaveragedQuantumDynamics2010, faisalTheoryMultiphotonProcesses1987, macriCoarseGrainedEffectiveHamiltonian2023}. A general recipe for the inclusion of time-dependence, both envelopes and chirps, in effective Hamiltonians beyond adiabatic elimination is unknown.%; the case of envelopes has has been studied by Faisal \cite{faisalTheoryMultiphotonProcesses1987}, with little to no adoption.

In this paper, we show that the open questions of the AMP approximation can be resolved by applying quasi-degenerate Rayleigh-Schrödinger perturbation theory (QD-RSPT) \cite{lindgrenAtomicManyBodyTheory1982}. This approach provides a clear pathway to obtain effective Hamiltonians for nearly-degenerate systems. The quasi-degenerate nature of QD-RSPT enables us to consistently describe detunings; this we show to resolve the debate on the choice of the interaction picture. The corrections to the AMP approximation lead to the non-orthogonality of the eigenstates of the effective Hamiltonian. We will show that the non-orthogonality expresses itself through the transient population of non-resonant states. In this way, ambiguities of the AMP are clarified. %If the non-resonant states can be reached via both photon emission and photon absorption from an essential state, 

Our perspective is naturally limited. Our focus on QD-RSPT implies that we ignore the significant body of work on time-averaged effective Hamiltonians. These averaging procedures are either based on the Dyson expansion \cite{jamesEffectiveHamiltonianTheory2007, gamelTimeaveragedQuantumDynamics2010}, or the Magnus expansion \cite{faisalTheoryMultiphotonProcesses1987, macriCoarseGrainedEffectiveHamiltonian2023} and have recently been utilized for the description of counter-rotating oscillations in quantum control theory \cite{barajasMultiTimescaleCoherentControl2025, barajasQuantumAveragingTheory2025}. Furthermore, the Schrieffer-Wolff transformation \cite{schriefferRelationAndersonKondo1966} yields an effective Hamiltonian in terms of operator corrections instead of energy corrections, which is particularly useful in quantum optics. Other names for the Schrieffer-Wolff transformation are van Vleck perturbation theory \cite{shavittQuasidegeneratePerturbationTheories1980}, or block diagonalization \cite{cederbaumBlockDiagonalisationHermitian1989}; related methods are Klimov's method of small rotations \cite{klimovMethodSmallRotations2000, klimovEffectiveHamiltoniansQuantum2002}, and the Morris-Shore transformation \cite{morrisReductionDegenerateTwolevel1983, torosovAdiabaticEliminationNearly2012, zlatanovMorrisShoreTransformationNondegenerate2020}, see also Ref.~\cite{sanzAdiabaticEliminationEffective2016, huangTheoryMultiphotonProcesses2025}. %Recently, this approach was combined with Floquet theory in the context of quantum control \cite{huangTheoryMultiphotonProcesses2025}. % in order to describe the full dynamics of few-level systems, both slow (Rabi oscillations) and fast (counter-rotating oscillations) \cite{huangTheoryMultiphotonProcesses2025} in a manner related to Ref.~\cite{bruhnkeGiantCounterrotatingOscillations2025}. %While we will comment on the connections between our perspective and the unitary-transformation perspective, we will not discuss the latter in any detail. 
%For a recent work on the connections between the time-independent effective Hamiltonians in the context of light-matter interaction, we refer the reader to Sanz \textit{et al.} (2016) \cite{sanzAdiabaticEliminationEffective2016}. 
A further limitation is that QD-RSPT is a time-independent theory and thus only applies to time-independent light-matter Hamiltonians. As soon as the laser field is shaped temporally by an envelope, or the laser pulse is chirped, we require a time-dependent approach. For effective Hamiltonians obtained via adiabatic elimination, envelopes are convenient to incorporate %, since the $n$th term of the effective Hamiltonian is proportional to the $n$th power of the field strength 
\cite{hansonManifestationsAtomicCore1997, vitanovPopulationTransferDecaying1997, yatsenkoTwophotonExcitationMetastable2005, tothProbingStrongfieldTwophoton2021, zhangEffectNonresonantStates2022, tothRoleDynamicStark2023}. We are only aware of one proposal that extends this approach to perturbative effective Hamiltonians beyond adiabatic elimination \cite{faisalTheoryMultiphotonProcesses1987}, and to our best knowledge, it was never applied. Incorporating chirp is even more elusive and will be addressed in upcoming work. %The second part of our work will address this time-dependent case.

Our work is structured as follows. We start with an elaborate account of QD-RSPT, in which we introduce the \textit{reduced wave operator} as the central object in perturbation theory. We further discuss the straight-forward extension for complex-scaled Hamiltonians, which allow us to treat ionization through a monotonous decrease in norm. In the results, we first compare the usual effective Hamiltonian from adiabatic elimination to that obtained through QD-RSPT. This reveals both the physical significance of the non-orthogonality of eigenstates and a non-degenerate model space. We provide three examples where the AMP approximation fails to capture physical properties: a toy model system, Rubidium in an IR field, and helium in an XUV field. Our effective Hamiltonian calculations are supported by Floquet calculations \cite{chuRecentDevelopmentsSemiclassical1985}, where the atomic parameters of Rubidium are obtained through a single-active electron (SAE) potential \cite{schweizerMODELPOTENTIALSALKALI1999}, and those of helium obtained through many-body configuration-interaction singles calculations \cite{foresmanSystematicMolecularOrbital1992, dreuwSingleReferenceInitioMethods2005}. Atomic units are used unless otherwise stated: $e = \hbar =  m = 1/4\pi\epsilon_0 = 1$.

\section{II. Theory}

Our starting point is the time-independent light-matter Hamiltonian
\begin{equation}
 H = H_\mathrm{atom} + H_\mathrm{field} + V \equiv H_0 + V.   
\end{equation}
For this section, it is inconsequential at what level of theory $H$ is obtained, be it quantum optics, Floquet theory, or simply within the rotating-wave approximation in the rotating frame. We assume the uncoupled eigenvalue problem $H_0 \ket{n} = E_n \ket{n}$ to be solved and write $H$ in the eigenbasis of $H_0$. The eigenvalues of $H$ we refer to as the dressed energies, and the eigenstates the dressed states. Let us for now assume that $H$ is Hermitian, and has a discrete spectrum.

In the upcoming sections, we will first derive the effective Hamiltonian as a formal object. Then, we derive the perturbative expansion for the effective Hamiltonian within QD-RSPT. We then elaborate on Floquet theory, which offers a pathway to obtain a time-independent light-matter Hamiltonian. Finally, we show the required adjustments that allow us to describe ionization through complex-scaling techniques. In our account, we stay close to the approach detailed by Lindgren and Morrison for the atomic many-body problem \cite{lindgrenRayleighSchrodingerPerturbationLinkeddiagram1974, lindgrenAtomicManyBodyTheory1982}, but with electron-electron correlation being replaced by the light-matter interaction.

\subsection{A. What is an effective Hamiltonian?}

An effective Hamiltonian is an operator that reproduces a subset of the true eigenvalues of the full Hamiltonian $H$. It acts not in the full Hilbert space, but rather in a Hilbert subspace $\P$, referred to as the model space, where $p \coloneqq \dim(\P)$. The model space is spanned by the essential states. We choose the essential states as those that the electric field couples (near-) resonantly from the ground state (GS). All other states are referred to as nonessential. The orthogonal space, spanned by nonessential states, is denoted by $\Q = P^\perp$ so that $\P\oplus\Q$ is the full Hilbert space. We define the projectors onto $\P$ and $\Q$ via
\begin{align}
    P = \sum_{n \in \P} \ketbra{n}{n} \ \text{ and } \ Q = 1 -P,
\end{align}
respectively. From this construction, it follows that $[H_0, P] = [H_0, Q] = 0$, as well as $PQ = QP = 0$, $P^2 = P$, $Q^2 = Q$, $PHQ = PVQ$, and $QHP = QVP$.

\subsection{B. Formal definition of the effective Hamiltonian}

We assume that $p$ eigenstates $\ket{\Psi_k}$ ($k = 1,...,p$) of the full Hamiltonian, $H\ket{\Psi_k} = \lambda_k\ket{\Psi_k}$, have their major part within the model space. Their projections onto $\P$ are
\begin{equation}
    \ket{\psi_k} = P\ket{\Psi_k}, \ \ k=1,...,p.
\end{equation}
We now define the wave operator $\wave$, denoted with a hat in order to distinguish it from the Rabi frequency $\Omega$, which reverses the above projection:
\begin{equation} \label{eq:waveoperator}
    \ket{\Psi_k} = \wave \ket{\psi_k}, \ \ k=1,...,p.
\end{equation}
The wave operator $\wave$, when acting on any eigenstate projection $\ket{\psi_k}$, $k=1,...,p$, yields the respective full eigenstate $\ket{\Psi_k}$ of $H$. We can rewrite $\wave = P + \chi$, where $\chi = Q\chi P$ is the reduced wave operator, also called correlation operator, which upon application to a $\P$-space state yields the corresponding components in the complementary space $\Q$, i.e., 
\begin{equation} \label{eq:reducedwaveoperator}
    Q\ket{\Psi_k} = \chi \ket{\psi_k}, \ \ k=1,...,p.
\end{equation}
Notably, $\chi$ is nilpotent, $\chi^2 = 0$. By applying $P$ to the time-independent Schrödinger equation (TISE) from the left,
\begin{equation}
    PH \wave \ket{\psi_k} = E_k \ket{\psi_k}, \ \ k=1,...,p,
\end{equation}
we can identify
\begin{equation} \label{eq:Heff2}
    H_\eff = PH\wave = PH(P+\chi) = PH_0P + PVP + PV\chi
\end{equation}
as the energy-independent effective Hamiltonian with eigenfunctions $\ket{\psi_k}$ and eigenvalues $E_k$ ($k=1,...,p$). Hence, the task of finding $H_\eff$ is formally reduced to finding the reduced wave operator. In this work, the mathematical properties of $\chi$ (existence, uniqueness) are assumed; we refer to Refs.~\cite{durandDirectDeterminationEffective1983, brouderRayleighSchrodingerPerturbationSeries2012, sanzAdiabaticEliminationEffective2016} for more information.

% Although not of practical importance to us, we will briefly comment on existence and uniqueness of $\chi$. Existence can be guaranteed under the requirement that all $P\ket{\Psi_k}$, $k=1,...,p$ are linearly independent \cite{durandDirectDeterminationEffective1983}. Uniqueness of $\chi$ cannot be guaranteed. However, the defining equation for $\chi$ is a Sylvester equation, which has a unique solution if the spectra of $PHP$ and $QHQ$ are disjoint \cite{brouderRayleighSchrodingerPerturbationSeries2012, sanzAdiabaticEliminationEffective2016}. If a $\Q$-space state has an eigenvalue that lies inside the $\P$-space spectrum, this condition is violated and we speak of an \textit{intruder state}. The intruder state problem has long vexed nuclear structure physicists \cite{hofmannNonperturbativeApproximationSchemes1974} as well as atomic and molecular structure theorists \cite{witekIntruderStateAvoidance2002}. We shall not concern ourselves with these troubles here. In some situations, the intruder state problem is trivially solved by including the intruder state in the $\P$-space (see our upcoming example for Rabi oscillations in rubidium). In other situations, perturbation theory breaks down spectacularly, and with it the assumption that our system is driven only by a handful of dressed energies. Then, exact methods such as Floquet theory \cite{chuFloquetTheoremGeneralized2004}, R-Matrix calculations \cite{burkeRMatrixTheoryAtomic2011} or close-coupling approaches \cite{burkeElasticScatteringLowEnergy1962} become indispensable.

While $H$ is Hermitian, $H_\eff$ is not. This is quite intuitive: $H$ has eigenvectors $\ket{\Psi_k}$, which are orthogonal. Meanwhile, $H_\eff$ has the eigenstates $\ket{\psi_k}= P\ket{\Psi_k}$, which by construction are not orthogonal. Hence, $H_\eff$ must generally be non-Hermitian, although it still has real eigenvalues $\lambda_k$. Another way to see this is to realize that due to $\chi^2 = 0$, $H_\eff$ can be expressed as a similarity transform of $H$,
\begin{equation} \label{eq:Heff_similarity_with_chi}
    H_\eff = P \e^{-\chi} H \e^{\chi} P.
\end{equation}
Since $\e^\chi = 1+\chi$ is not unitary, $H_\eff$ is generally non-Hermitian \cite{suzukiDegeneratePerturbationTheory1983}. 

The choice of basis in quantum mechanics is of course inconsequential to the observables. Therefore, it is possible to find an Hermitian effective Hamiltonian with the same eigenvalues $\lambda_k$ as $H_\eff$ from Eq.~\eqref{eq:Heff2}. To construct this Hermitian effective Hamiltonian, which we denote via $\H_\eff$, we notice that the properties of $\chi$, and orthogonality of the full eigenstates of $H$ yield
\begin{multline}
        \delta_{nm} = \braket{\Psi_n|\Psi_m} = \braket{\Psi_n|P|\Psi_m} + \braket{\Psi_n | Q |\Psi_m} \\ = \braket{\psi_n| P + \chi^\dagger\chi |\psi_m}
\end{multline}
Note that $P+\chi^\dagger\chi$ is Hermitian. This suggests that the $\P$-space eigenfunctions defined by
\begin{equation}
    \ket{\phi_k} = (P+\chi^\dagger\chi)^{1/2} \ket{\psi_k} \eqqcolon \eta\ket{\psi_k} \text{ and } \bra{\phi_k}=\bra{\psi_k}\eta
\end{equation}
are orthogonal. Hence, 
\begin{align} \label{eq:heffhermitianexact}
     \H_\eff = \eta H_\eff \eta^{-1}
\end{align}
must be Hermitian, with orthogonal eigenstates $\ket{\phi_k}$. Note that $ \H_\eff$ is not ``manifestly Hermitian'', which is a common phrase used to express that its Hermiticity cannot be seen by eye \cite{brandowFormalTheoryEffective1979, suzukiDegeneratePerturbationTheory1983, killingbeckBlochWaveOperator2003}, but that it must instead be proven explicitly, see for instance Ref.~\cite{sanzAdiabaticEliminationEffective2016}. The implication of this is profound: $ \H_\eff$ is only exactly Hermitian if $\chi$ is the exact reduced wave operator and $H_\eff$ is the exact effective Hamiltonian. If either $H_\eff$ or $\chi$ are obtained approximately, $\H_\eff$ will only be approximately Hermitian. 
%
%\textcolor{blue}{There exist deep connections to non-Hermitian quantum theory, in which non-Hermitian Hamiltonians with real eigenvalues are equivalently referred to as pseudo-, quasi-, or crypto-Hermitian; they are then related via similarity transform to a Hermitian Hamiltonian. Since the questions of ``physicality'' }

%they generate the same dynamics. This is easily demonstrated. Let $\ket{\psi(t)} \in \P$. We start with the TDSE for $H_\eff$,
%
%\begin{equation}
%     \i \ddt \ket{\psi(t)} = H_\eff \ket{\psi(t)}.
%\end{equation}
%
%Using $\ket{\phi(t)} = \eta\ket{\psi(t)}$ and multiplying with $\eta$ from the left, we obtain
%
%\begin{equation}
%    \i \eta \ddt \ket{\psi(t)}   = \eta H_\eff \eta^{-1} \ket{\phi(t)} 
%\end{equation}
%
%and directly identify that
%
%\begin{equation}
%        \i \ddt \ket{\phi(t)} = \H_\eff \ket{\phi(t)}.
%\end{equation}

For context, we lastly note that $\H_\eff$ is the object obtained via the Schrieffer-Wolff transformation and its related variants \cite{shavittQuasidegeneratePerturbationTheories1980, cederbaumBlockDiagonalisationHermitian1989, morrisReductionDegenerateTwolevel1983, zlatanovMorrisShoreTransformationNondegenerate2020, klimovMethodSmallRotations2000}. In the context of light-matter interaction, we want to highlight in particular the recent work of Sanz \textit{et al.} \cite{sanzAdiabaticEliminationEffective2016}, who treat adiabatic elimination as a singular perturbation problem, with the reduced wave operator $\chi$ (in their work denoted $B$) as the central object, connecting both $H_\eff$ and $\H_\eff$. Even more recently, Huang \textit{et al.} employed the effective Hamiltonian $\H_\eff$ in quantum control problems, focusing among others on the fast dynamics due to counter-rotating oscillations in few-level systems \cite{huangTheoryMultiphotonProcesses2025}.

\subsection{C. Quasi-degenerate Rayleigh-Schrödinger perturbation theory}

QD-RSPT offers the possibility to calculate effective Hamiltonians beyond the pole approximation in a perturbative manner. The quasi-degeneracy refers to the model space: $PH_0P \neq E_0P$. This is the case in light-matter interaction when the light is not exactly resonant, but slightly detuned. We will show that it is a strength of QS-RSPT that the description of detuning is natural in the formalism, while the AMP effective Hamiltonian is not suited for non-negligible detunings. %Perhaps surprisingly, QD-RSPT has had little adoption in the calculation of effective Hamiltonians in light-matter interaction. While in the 1970s, the resolvent technique was popular \cite{lambropoulosTheoryMultiphotonIonization1974}, and closely related to it, Brillouin-Wigner perturbation theory approaches in conjunction with continuous fraction expansions \cite{gontierExactSummationHigherorder1976, jacksonComparisonProjectionOperator1981}, the most popular approach today is arguably adiabatic elimination \cite{shoreTheoryCoherentAtomic1990}, which has been shown to relate closely to both the resolvent technique in the pole approximation \cite{brionAdiabaticEliminationLambda2007}, and the Markov approximation \cite{paulischAdiabaticEliminationHierarchy2014}. %There exist a multitude of ways to arrive at the RS expansion, several of them differing further in the manner in which the non-degeneracy is treated. We will present the approach by Lindgren (1974) \cite{lindgrenRayleighSchrodingerPerturbationLinkeddiagram1974, lindgrenAtomicManyBodyTheory1982}.%, since we will show in a later section how it elegantly generalizes to slowly-varying time-dependent perturbations (e.g. for the description of pulse envelopes). 

In the previous section, we showed that the problem of finding the effective Hamiltonian can be reformulated to the problem of finding the reduced wave operator. With the definitions $Q\ket{\Psi} = \chi P \ket{\Psi}$, and $\chi = Q\chi P$, we find immediately that the reduced wave operator fulfills the \textit{decoupling equation} \cite{suzukiDegeneratePerturbationTheory1983}, 
\begin{align}
    0 &= Q(\e^{-\chi} H \e^\chi)P =  Q(1-\chi)H(1+\chi)P, \nonumber \\
    &= QVP + QH\chi  - \chi HP - \chi V\chi .
\end{align}
For a proof, let $Q(1-\chi)H(1+\chi)P$ act on an arbitrary eigenstate $\ket{\Psi_k}$ of the full Hamiltonian. 

Simple rearrangements yield the generalized Bloch equation (GBE),
\begin{equation} \label{eq:blocheq}
    [\chi, H_0]  = QV\wave  - \chi V\wave .
\end{equation}

We now construct a perturbative series in $\wave$, and thus $\chi$:
\begin{equation} \label{eq:chipert}
    \wave = P+\chi = P + \chi^{[1]} + \chi^{[2]} + ...
\end{equation}
In the context of light-matter interaction, $\chi^{[n]}$ contains all processes in which $n$ photons are exchanged in order to get from a state $\ket{p}\in\P$ to a state $\ket{q}\in \Q$. We insert Eq.~\eqref{eq:chipert} into Eq.~\eqref{eq:blocheq}, and equate terms of the same order in $V$
\begin{align}
    [\chi^{[1]}, H_0]  &= QVP \\
    [\chi^{[2]}, H_0]  &= QV\chi^{[1]} - \chi^{[1]} VP \\
    &\vdots \nonumber \\
    [\chi^{[n]}, H_0]  &= QV\chi^{[n-1]} - \sum_{m=1}^{n-1}\chi^{[n-m]} V\chi^{[m-1]}
\end{align}
We obtain an explicit formula for $\chi^{[n]}$ by projecting from left with a $\Q$-space state $\bra{q}$, and from the right with a $\P$-space state $\ket{p}$. Defining $\chi_{qp}^{[n]} \coloneqq \braket{q|\chi^{[n]}|p} $, we obtain
\begin{multline}
    \chi_{qp}^{[n]} = \frac{1}{E_p - E_q} \bra{q} \biggl( QV\chi^{[n-1]} \\ - \sum_{m=1}^{n-1}\chi^{[n-m]} V\chi^{[m-1]} \biggr) \ket{p}.
\end{multline}
For notational convenience, we define element-wise multiplication in the atom-photon state basis via $\odot$ and write the resolvent matrix $G$ via its elements
\begin{equation} \label{eq:resolventRSPT}
    G_{qp} = \frac{1}{E_p - E_q},
\end{equation}
after which we can write
\begin{equation} \label{eq:waveoperator_general}
    \chi^{[n]} = G \odot \left( QV\chi^{[n-1]} - \sum_{m=1}^{n-1}\chi^{[n-m]} V\chi^{[m-1]} \right) .
\end{equation}
The effective Hamiltonian of RSPT of $n$th order therefore reads
\begin{equation} \label{eq:rspt}
    H_\RS^{(n)} = PH_0P + PVP + \sum_{k=1}^{n-1} PV \chi^{[k]},
\end{equation}
where the term $PV\chi^{[k-1]} \propto \efield_0^k$. Note that the convergence of the RS perturbative series for $n\rar\infty$ cannot be taken for granted, but robust results will often be achieved with truncation at a low orders \cite{benderAnharmonicOscillatorII1973, vrscayContinuedFractionsRayleigh1986, jentschuraResummationDivergentPerturbation2001}.

If the model space is degenerate, i.e. $PH_0P = E_0P$, the expressions simplify to
\begin{equation} \label{eq:degenerateRS_chi}
    \chi^{[n]} = G_\Q \left(QV\chi^{[n-1]} - \sum_{m=1}^{n-1}\chi^{[n-m]} V\chi^{[m-1]}\right),
\end{equation}
where we defined the $\Q$-space resolvent
\begin{equation} \label{eq:qspaceresolvent}
    G_\Q \coloneqq \frac{Q}{E_0 - QH_0Q}.
\end{equation}

It should be noted here that while the matrix elements of $H_\RS^{(n)}$ can of course be written down with pen and paper, the number of terms quickly become prohibitive for $n\geq 4$. The utility of QD-RSPT in light-matter interaction comes from the ease and speed at which it can be calculated in high-level programming languages such as Python. Furthermore, since the $k$-th term of $H_\RS$ is proportional to $\efield_0^k$, we can obtain $H_\RS$ at any field strength $\efield_0$ by calculating the perturbative expansion of $H_\RS$ at a single $\efield_0$ once (the limit being of course the convergence of $H_\RS$ at high $\efield_0$). In this way, QD-RSPT provides a generalization to the perturbative AMP effective Hamiltonian, where scaling parameters have previously been provided \cite{olofssonPhotoelectronSignatureDressedatom2023}.

\subsection{D. Floquet theory}

So far, we have left the light-matter Hamiltonian $H$ unspecified. In this section, we offer a brief overview of Floquet theory, which lets us obtain a time-independent, semi-classical light-matter Hamiltonian $H_F$. This Hamiltonian can be obtained from the semiclassical, periodic $H(t) = H_0^\mathrm{(atom)} + z E_0 \cos(\omega t)$ via Fourier techniques. Here, $E_0$ is the electric field amplitude and $z$ the dipole operator with matrix elements $z_{ij}$. Alternatively, the quantum-optics Hamiltonian can be transformed into $H_F$ for a coherent photon field in the limit of infinite cavity volume and infinite photon numbers \cite{shirleySolutionSchrodingerEquation1965, guerinRelationCavitydressedStates1997}. These assumptions tend to be valid in intense fields. Recently, strong-field physics and quantum optics have begun to merge \cite{lewensteinGenerationOpticalSchrodinger2021, stammerQuantumElectrodynamicsIntense2023, tsatrafyllisHighorderHarmonicsMeasured2017, gorlachQuantumopticalNatureHigh2020, gothelfHighorderHarmonicGeneration2025, stenquistEntanglementTransferComposite2025}, an exciting development which lies outside the scope of this work. For reviews and book chapters on Floquet theory, we refer the reader to Ref.~\cite{chuFloquetTheoremGeneralized2004, joachainAtomsIntenseLaser2011}. While we will write down the equations for linearly polarized light, Floquet theory can be elegantly formulated just as well for circularly polarized light \cite{chuQuasienergyFormalismIntense1978}. Further, many-mode Floquet theory allows for the convenient description of bichromatic pulses \cite{hoSemiclassicalManymodeFloquet1983}.

We choose the semi-classical Hamiltonian $H(t)$ as our starting point. The Floquet theorem permits us to expand the time-dependent wavefunction in the form
\begin{equation}
    \ket{\Psi^{(\lambda)}(t)} = \e^{-\i \lambda t} \sum_{k=-\infty}^\infty \e^{\i k \omega t} \ket{\phi_k^{(\lambda)}}, \label{eq:wavefunctionfloquet}
\end{equation}
labelled by the quasi-energy $\lambda$. Upon substitution into the TDSE, 
\begin{equation}
    \i \ddt \ket{\Psi(t)} = H(t) \ket{\Psi(t)},
\end{equation}
we can group the resulting equations by the harmonic component to obtain the set of equations
\begin{equation}
    (H_0^\mathrm{(atom)} + k \omega) \ket{\phi_k^{(\lambda)}} + \frac{\efield_0 z}{2} \left( \ket{\phi_{k+1}^{(\lambda)}} + \ket{\phi_{k-1}^{(\lambda)}} \right) = \lambda \ket{\phi_k^{(\lambda)}},
\end{equation}
where $k$ runs over all negative and positive indices. The harmonic vectors $\ket{\phi_k^{(\lambda)}}$ are then collected as a column vector $\ket{\phi^{(\lambda)}}$, known as the quasi-energy state, so that we obtain the simple eigenvalue equation
\begin{equation}
    H_F \ket{\phi^{(\lambda)}} = \lambda \ket{\phi^{(\lambda)}},
\end{equation}
where $\lambda$ is an eigenvalue of $H_F$. The eigenvectors and -values are often referred to as dressed states and dressed energies due to their close connection with quantum optics. The Floquet Hamiltonian can be written in an extended Hilbert space (known as the Sambe space) via
\begin{equation} \label{eq:floquethamil}
    H_F = H_0^\mathrm{(atom)} \otimes \mathbbm{1} + \omega \mathbbm{1} \otimes D + \frac{\efield_0}{2} z \otimes F,
\end{equation}
where $D = \mathrm{diag}(...,-k, ..., 0, ... ,k',...)$, and $F$ has ones on the sub- and superdiagonals ($F_{ij} = \delta_{i+1,j} + \delta_{i,j+1}$). In perturbation theory, we will then take $H_0^\mathrm{(atom)} \otimes \mathbbm{1} + \omega \mathbbm{1} \otimes D$ as the unperturbed Hamiltonian $H_0$ whose solution is known, and $\frac{\efield_0}{2} z \otimes F$ as the perturbation $V$. The unperturbed states are defined as $\ket{n,k} = \ket{n}\otimes \ket{k}$, with $H_0^\mathrm{(atom)}\ket{n} = E_n\ket{n}$, and $D \ket{k} = k \ket{k}$. In analogy to quantum optics, we call them the (uncoupled) atom-photon states, which should however not cover up the fact that Floquet theory is a semi-classical theory. 

%From Eq.~\eqref{eq:wavefunctionfloquet}, we can deduce that the transition amplitude between two atomic states $\ket{a}$ and $\ket{b}$ can be expressed as a coherent sum of the transition amplitudes between atom-photon states with varying harmonic index $k$, weighted by the harmonics $\exp(\i k\omega t)$:
%
%\begin{align} \label{eq:transitionamplitude_floquet}
%    U_{ba}(t,t_0) = \sum_k \braket{b,k | \e^{-\i H_F (t-t_0)} | a,0} \e^{\i k\omega t}.
%\end{align}
%
%The other amplitudes $U_{aa}(t,t_0)$, etc., follow accordingly. %Eq.\eqref{eq:transitionamplitude_floquet} forms the basis for our description of counter-rotating oscillations in Section~V.

%We finally note that the concept of quasi-energies can also be conveniently applied to circular polarization \cite{chuQuasienergyFormalismIntense1978}. For right-hand circularly polarized light in the $x$-$y$-plane, the semi-classical, time-dependent light-matter Hamiltonian can be rotated into the co-rotating frame, yielding
%
%\begin{equation}
%    H_F = H_0^{(\mathrm{atom})} - \omega L_z + \frac{\efield_0}{\sqrt{2}} x,
%\end{equation}
%
%where $L_z$ is the $z$-component of the angular momentum operator, and $x$ is the dipole operator along the $x$-axis. 

\subsection{E. Complex-scaling}

In the form proposed by Shirley (1965), Floquet theory was only applicable to bound states \cite{shirleySolutionSchrodingerEquation1965}. In the following decades, Floquet theory was extended to also allow for the treatment of the continuum via complex-scaling techniques \cite{chuIntenseFieldMultiphoton1977, maquetStarkIonizationDc1983}. These techniques rotate the spectrum of $H_F$ into the complex plane by an angle $\theta$ so that the continuum may be discretized by expanding into an $L^2$-basis. Often, the transform is only applied beyond a certain radius, which is known as exterior complex-scaling \cite{simonDefinitionMolecularResonance1979}. The Floquet Hamiltonian of Eq.~\eqref{eq:floquethamil} then becomes complex-symmetric with eigenvalues $E - \frac{\i}{2}\Gamma$, where $E\in\mathbb{R}$ is the dressed energy, and $\Gamma \in \mathbb{R}$ is the decay rate (equivalently, the width) of the state. 

Traditionally, the complex-scaled Floquet Hamiltonian $H_F^\theta$ was constructed via expansion into Sturmian functions \cite{potvliegeTimeindependentTheoryMultiphoton1988}, or Laguerre functions \cite{chuIntenseFieldMultiphoton1977}. Another approach is to solve the TISE with complex-scaling in another suitable basis, such as B-Splines, and subsequently construct $H_F^\theta$ from the complex-scaled atomic parameters \cite{bruhnkeGiantCounterrotatingOscillations2025}. 

Dealing with a complex-symmetric Hamiltonian $H$ necessitates dealing with non-Hermitian quantum theory \cite{moiseyevNonHermitianQuantumMechanics2011}. Under the usual scalar product $\braket{\: \cdot \: | \: \cdot \: }$, the adjoint $\bra{ \: \cdot \:}$ of the ket vector $\ket{\: \cdot \:}$ is given by $\bra{\: \cdot \:} = (\ket{\: \cdot \:})^\dagger$. Clearly, the eigenvectors of a complex-symmetric $H$ are not orthogonal under this scalar product. To recover the concept of orthogonality for complex-symmetric Hamiltonians, we define the bra-vector via the transpose $\bra{ \: \cdot \:} \coloneqq (\ket{ \: \cdot \:})^T$. Then, if $\ket{n}$ and $\ket{m}$ are eigenfunctions of a complex-symmetric $H$ with $H \ket{n} = \lambda_n \ket{n}$, $\lambda_n \in \mathbb{C}$, we have $\braket{n|m} = \braket{m|n}$ and $\braket{n|H|m} = \braket{m|H|n}$, leading to the orthogonality relation \cite{holtTimeDependencesTwo1983}
\begin{equation}
    (\lambda_n - \lambda_{m}) \braket{n|m} = 0,
\end{equation}
which guarantees the orthogonality of the eigenvectors of a symmetric Hamiltonian. Further, we obtain the usual closure identity $1 = \sum_n \ketbra{n}{n}$. We stress that these considerations only apply for a complex-symmetric $H$. For more general non-Hermitian $H$, the c-product with soft brackets $( \: \cdot \:  | \: \cdot \: )$ needs to be introduced, which is based on the concept of left and right eigenvectors and implies that the complex-conjugation of the bra-vector is performed only on those parts of the wavefunction that would be complex \textit{without} complex-scaling \cite{moiseyevNonHermitianQuantumMechanics2011}. Complex-symmetric $H$ are obtained if the light is linearly polarized along $z$, and also in circular polarization if the Hamiltonian is transformed into a suitable rotating frame \cite{chuQuasienergyFormalismIntense1978}. We can conveniently adapt QD-RSPT to obtain a complex-scaled $H_\eff$ from $H_F^\theta$ \cite{joachainAtomsIntenseLaser2011}.

Clearly, we have to justify this introduction of ambiguous notation, in which the Dirac brackets are re-used for the modified scalar product. The advantage lies in the ability to generalize the following theoretical explorations. All upcoming formulas may be interpreted either within Hermitian or non-Hermitian quantum theory, simply by exchanging Hermitian conjugates and transposes, both for operators and for bra-vectors. For example, for Hermitian $H$, the norm of the wavefunction $\ket{\Psi}$ can be expressed through the $\P$-space wavefunction $\ket{\psi} \coloneqq P \ket{\Psi}$ as
\begin{equation}
    \braket{\Psi|\Psi} = \braket{\psi|P+\chi^\dagger\chi|\psi}.
\end{equation}
For complex-symmetric $H$, it is given by
\begin{equation}
    \braket{\Psi|\Psi} = \braket{\psi|P+\chi^T\chi|\psi},
\end{equation}
where the use of the $c$-product was implied. 

Recall that for Hermitian $H$, the effective Hamiltonian $H_\eff$ is generally non-Hermitian, while $\H_\eff$ is Hermitian. Analogously, for complex-symmetric $H$, the effective Hamiltonian $H_\eff$ is generally complex non-symmetric, while $\H_\eff$ is complex-symmetric.

%Lastly, if the Hamiltonian $H$ is neither Hermitian nor complex-symmetric, biorthogonality can still be fulfilled by inspecting the left and right eigenvectors of $H$:
%
%\begin{align}
%    H\ket{n} &= \lambda_n \ket{n}, \\
%    \bra{\tilde n} H &= \bra{\tilde n} \lambda_n.
%\end{align}
%
%We refer to the left eigenvector $\bra{\tilde n}$ as the biorthogonal state of $\ket{n}$. Then, we can normalize the states so that the biorthogonality relation
%
%\begin{align}
%    \braket{\tilde n| n'} = \delta_{nn'}
%\end{align}
%
%holds, and the closure relation reads $\sum_n \ketbra{n}{\tilde n} = 1$. This notation will make it convenient to distinguish between Hamiltonians where both $H_{ij} \neq H_{ji}^\ast$ (Hermiticity) and $H_{ij} \neq H_{ji}$ (complex symmetry). When we speak in the following sections of non-orthogonal eigenstates of a Hamiltonian, it means that the operator is neither Hermitian, nor complex-symmetric, thus necessitating the use of the biorthogonal complement.

\section{III. Results}

\subsection{A. Relation to the AMP approximation and the role of non-orthogonal eigenstates}

%Adiabatic elimination (AE) and the pole approximation (PA) are two ways to obtain an equivalent effective Hamiltonian \cite{brionAdiabaticEliminationLambda2007}. 
In this section, we compare the effective Hamiltonian from the AMP approximation to that of QD-RSPT, and propose a resolution to a longstanding debate on the proper choice of the interaction picture for the AMP approximation. Note that our expressions in this section are for Hermitian Hamiltonians $H = H_0 + V$. They can be easily reformulated for complex-symmetric Hamiltonians.

The AMP effective Hamiltonian can be formulated as
\begin{align} \label{eq:Heff_AE}
    H_\Ae = PHP + PV\frac{Q}{E_0 - QHQ} VP;
\end{align}
we refer to Appendix~A for derivations using adiabatic elimination, the Markov approximation, and the pole approximation. The energy $E_0$ is in principle a free parameter. However, it is clear from the AMP approximation that $E_0$ should be chosen to lie in the vicinity of the $\P$-space spectrum. For sufficiently weak interactions, we can expand the inverse $[E_0 - QHQ]^{-1}$ in the interaction,
\begin{multline}
    \frac{Q}{E_0 - QHQ} = G_\Q + G_\Q V G_\Q + G_\Q V G_\Q V G_\Q + ...,
\end{multline}
so that the AMP effective Hamiltonian in $n$th order perturbation theory reads
\begin{align} \label{eq:Heff_AE_pert}
    H_\Ae^{(n)} = PHP + \sum_{k=1}^{n-1} PV\left(G_\Q V\right)^kP,
\end{align}
from which we can see by eye that the corresponding reduced wave-operator is defined via the recursion
\begin{equation}\label{eq:degenerateAE_chi}
    \chi^{[n]} = G_\Q V\chi^{[n-1]},
\end{equation}
with $\chi^{[1]} = G_\Q VP$. Eq.~\eqref{eq:degenerateAE_chi} should be compared with the expression for $\chi^{[n]}$ from degenerate RSPT, Eq.~\eqref{eq:degenerateRS_chi}. The approximations of $H_\Ae$ enter through the neglect of $G_\Q \sum_{m=1}^{n-1} \chi^{[n-m]}V\chi{[m-1]}$.

Note how the effective interaction always follows the pattern
\begin{equation} \label{eq:pqqp_path}
    \P \rar \Q \rar \underbrace{ \ ... \ }_{\text{only via }\Q} \rar \Q \rar \P,
\end{equation}
in other words, the only allowed $\P$-space states are the initial and final state. These paths are necessarily always \textit{symmetric}, which provides the physical explanation as to why $H_\Ae$ is Hermitian. We deduce that the attribute of all intermediate states lying in $\Q$ is the defining attribute for the AMP effective Hamiltonian.

It is instructive to compare this perturbation series with the one that we showed for the degenerate RS expansion in Eq.~\eqref{eq:degenerateRS_chi}. First, we notice that in degenerate RSPT, $E_0$ is unambiguously defined as the unperturbed $\P$-space eigenenergy $PH_0P = E_0P$. The lowest-order correction to $H_\Ae$ comes in the third order of the effective Hamiltonian, i.e., in second order of the reduced wave-operator, $\chi^{[2]}$, where we subtract $G_\Q \chi^{[1]} VP$ from $G_\Q V\chi^{[1]}$. In the effective Hamiltonian, this term appears in third order as $-PV G_\Q \chi^{[1]} VP$. This is is the first term that causes the eigenstates of $H_\eff$ to be non-orthogonal. It represents a three-photon pathway of the kind $\P \rar \P \rar \Q \rar \P$, where a first-order coupling in the $\P$-space is composed together with a second-order coupling via the $\Q$-space (we read $-PV G_\Q \chi^{[1]} VP$ from right to left). For a two-level $\P$-space with degenerate states $\ket{a}$ and $\ket{b}$ (unperturbed energy $E_0$), the off-diagonals of the three-photon pathway read
\begin{align}
    (-PVG_\Q \chi^{[1]} VP)_{ba} &= -\sum_q \frac{V_{bq} V_{qb}}{(E_0 - E_q)^2} V_{ba}, \\
    (-PVG_\Q \chi^{[1]} VP)_{ab} &= -\sum_q \frac{V_{aq} V_{qa}}{(E_0 - E_q)^2} V_{ab} .
\end{align}
While $V_{ba} = V_{ab}^\dagger$, in general the two-photon couplings $V_{bq} V_{qb} \neq V_{aq} V_{qa}$ are not equal. This elucidates the origin of the non-Hermiticity. Only if $\ket{a}$ and $\ket{b}$ couple equally strongly to the $\Q$-space will $H_\RS$ be Hermitian. In general systems, the coupling from $\P$ to $\Q$ is asymmetric \cite{bruhnkeGiantCounterrotatingOscillations2025}. %Clearly, in QD-RSPT, the analogue object to adiabatic elimination is obtained by setting
%
%\begin{equation}
%    G \odot \sum_{m=1}^{n-1} \chi^{[n-m]} V \chi^{[m-1]} = 0.
%\end{equation}

\subsection{B. Quasi-degenerate AMP approximation}

In the literature on the AMP approximation, it has been discussed at length how to choose $E_0$, or equivalently, how to choose the correct interaction picture in terms of the energy off-set on the Hamiltonian's diagonal \cite{brionAdiabaticEliminationLambda2007, paulischAdiabaticEliminationHierarchy2014, macriCoarseGrainedEffectiveHamiltonian2023}. The most common proposal is to choose it as the center of the $\P$-space spectrum \cite{brionAdiabaticEliminationLambda2007}. To the best of our knowledge, it has not been proposed so far to just take the non-degeneracy into account properly, by calculating the AMP effective Hamiltonian in accordance to QD-RSPT as
\begin{align} \label{eq:Heff0_nondegenerate}
    H_\QDAE^{(n)} = PHP + \sum_{k=1}^{n-1} PV\left[G \odot (Q V)\right]^kP, 
\end{align}
where $G$ is defined as in Eq.~\eqref{eq:resolventRSPT} by the unperturbed $\P$-space eigenvalues. %This amounts to setting all photon pathways in QD-RSPT that are composed of lower-order photon pathways to zero,
%
%\begin{equation}
%    G \odot \sum_{m=1}^{n-1} \chi^{[n-m]} V \chi^{[m-1]} = 0.
%\end{equation}
%
The defining equation for the reduced wave-operator in this quasi-degenerate AMP approximation then becomes
\begin{equation}
    \chi^{[n]} = G \odot ( QV\chi^{[n-1]}),
\end{equation}
with $\chi^{[1]} = G \odot ( QVP)$, which may be compared to Eq.~\eqref{eq:degenerateAE_chi}. 

The effective Hamiltonian of Eq.~\eqref{eq:Heff0_nondegenerate} treats all essential states on an equal footing and therefore does not have an ambiguous parameter $E_0$. Crucially, it retains the defining property of the AMP approximation: All intermediate states are in the $\Q$-space. We provide in Section IV.B a realistic example for a strongly coupled rubidium atom, where the detuning, i.e., non-degeneracy, needs to be taken into account in order to accurately capture the physics.

We should note that $H_{\QDAE}$ from Eq.~\eqref{eq:Heff0_nondegenerate} is non-Hermitian despite only allowing paths via the $\Q$-space, i.e., those of Eq.~\eqref{eq:pqqp_path}. For example, in second order, we have
\begin{align}
    (H_{\QDAE}^{(2)})_{pp'} &= H_{pp'} + \sum_q \frac{V_{pq} V_{qp'}}{E_{p'} - E_{q}} \\
    (H_{\QDAE}^{(2)})_{p'p} &= H_{p'p} + \sum_q \frac{V_{p'q} V_{qp}}{E_{p} - E_{q}} ,
\end{align}
which means that $(H_{\QDAE})_{pp'} \neq (H_{\QDAE})^\dagger_{p'p}$ because $E_p \neq E_{p'}$. This non-Hermiticity expresses the most simple source of asymmetry in the coupling to the $\Q$-space: Since $\ket{p}$ and $\ket{p'}$ are non-degenerate, the detuning of $\ket{p}$ with any $\Q$-space state will be different than the detuning of $\ket{p'}$ with that same $\Q$-space state. Hence, the effective coupling must differ in general.

\subsection{C. Validity of the effective Hamiltonian approach}

In the well-explored V-type three-level system, the validity condition for adiabatic elimination is that the coupling $\Omega$ to the intermediate state should be much smaller than the detuning $\Delta$ of the intermediate state, $|\Omega/\Delta| \ll 1$ \cite{brionAdiabaticEliminationLambda2007}. When we have many intermediate states, the natural generalization is to search for the bound $\Q$-space state $\ket{q}$ that maximizes the ratio $|\Omega_{pq}/\Delta_{pq}|$ for any state $\ket{p}\in\P$ \cite{faisalTheoryMultiphotonProcesses1987}.

We can formalize this condition through the reduced wave operator $\chi$. For this, we notice that in QD-RSPT, the first order $\chi^{[1]}$ is defined via its matrix elements
\begin{equation}
    \chi_{qp}^{[1]} = \frac{V_{qp}}{E_p - E_q},
\end{equation}
in which $\Delta_{pq} \equiv E_p - E_q$ is the detuning of $\ket{q}$ with respect to $\ket{p}$. Hence, the formal version of the condition will be $\| \chi^{[1]}\|_{\max} \ll 1$. It should be noted that this condition applies only to the bound states. The continuum is of course always resonant, but does not cause singularities due to the imaginary part. Hence, only the projection of $\chi$ on the bound states should be considered. The condition then answers the question: How strongly at most can a $\P$-space state couple to a bound $\Q$-space state via one photon?

Clearly, the condition can trivially be generalized to answer the question of how strongly at most a $\P$-space state can couple to a bound $\Q$-space state via up to $N$ photons, $\left\| \sum_{n=1}^N \chi^{[n]} \right\|_{\max} \ll 1$, or, without specifying the order, 
\begin{equation}
\| \chi \|_{\max} \ll 1    .
\end{equation}
Thus, the reduced wave-operator, which couples the $\P$-space to the $\Q$-space, provides information on the validity and eventual breakdown of the effective Hamiltonian approach.

%Of course, other questions may also be posed and answered:  $\|\chi\|_1 = \max_p (\sum_q |\chi_{qp}|)$ predicts how strongly at most a $\P$-space state can couple to the entire $\Q$-space, while $\|\chi\|_\infty= \max_q (\sum_p |\chi_{qp}|)$ predicts how strongly at most a $\Q$-space state can couple to the entire $\P$-space. In lowest-order perturbation theory, these metrics scale with the same power of the field strength, so that any of them work as an indicator for the validity of QD-RSPT.

\section{IV. Examples}

With the following examples, we aim to illustrate three main points: (i) the physical implication of the non-orthogonal eigenstates of $H_\RS$, (ii) a system in which the non-degeneracy of the $\P$-space makes the AMP approximation fail at already low intensities, while the quasi-degenerate AMP approximation is reliable at those intensities, (iii) a system in which the eigenstates of $H_\RS$ are significantly non-orthogonal. To this end, we analyse Rabi oscillations in a simple model system in Section~IV.A, investigate rubidium in an IR field in Section~IV.B, and finally present results for helium strongly coupled by the XUV field in Section~IV.C. In these examples, we focus on the quasienergy structure of the systems, while the time-dependent implications are only discussed briefly in Section~IV.A to illustrate consequences of the non-orthogonal eigenstates of $H_\RS$. Time-dependent effective Hamiltonian calculations will be presented in upcoming work, where we will further include envelopes and chirps in a general formalism.

\subsection{A. Rabi oscillations in a model system}

\begin{figure*}[t]
    \centering
    \includegraphics{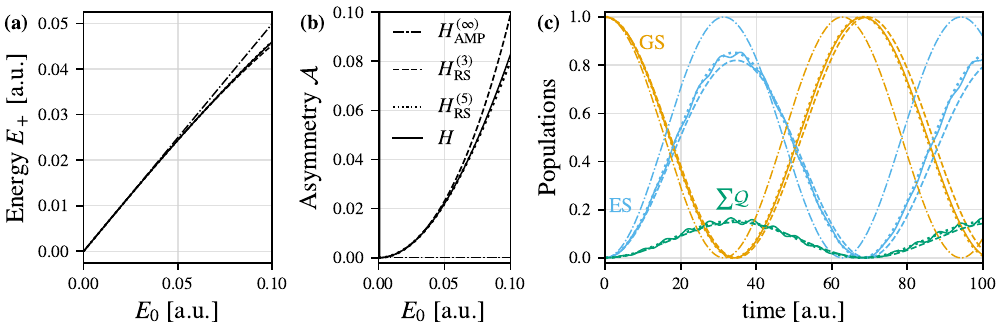}
    \caption{Properties of the minimal model Hamiltonian $H$ (solid) of Eq.~\eqref{eq:Hmodel}, compared with the effective Hamiltonians $H_\Ae^{(\infty)}$ (dashed-dotted), $H_\RS^{(3)}$ (dashed), and $H_\RS^{(5)}$ (dotted). Panel~(a) shows the dressed energy $E_+$ against field strength $E_0$. Panel~(b) depicts the effective asymmetry defined in Eq.~\eqref{eq:asymmetrymodel} against $E_0$. Panel~(c) shows the populations of the GS (yellow), ES (blue) and the sum of the $\Q$-space populations (green). }
    \label{fig:model_nonhermitian_H_eff}
\end{figure*}

It is now clear that $H_\eff$ should be non-Hermitian in general. In this example, we will demonstrate the implications, both for the dressed energies and the time-evolution, using a minimal model system. We consider three atomic levels: a GS $\ket{a}$, an excited state (ES) $\ket{b}$, and a third state $\ket{q}$. The GS with energy $E_a = 0$ is resonantly coupled to the ES with Rabi frequency $\Omega_{ab} = \Omega_{ba} = E_0 z_{ab}$. The frequency of the field is hence $\omega = E_b - E_a$. The third state $\ket{q}$ couples to the ES with $\Omega_{bq} = \Omega_{qb} = E_0 z_{qb}$; the latter two atomic states are furthermore degenerate, $E_q = E_b$. Thus, $\ket{b}$ couples to $\ket{q}$ non-resonantly, both via absorption and emission with equal strength. As such, the minimal relevant block of the Floquet Hamiltonian $H$ has four levels, where $\ket{a, N}$ and $\ket{b,N-1}$ make up the $\P$-space and $\ket{q, N - 1\pm 1}$ make up the $\Q$-space:
\begin{align} \label{eq:Hmodel}
    H =  \begin{pmatrix} 0 & \Omega_{ab}/2 & 0 & 0 \\\Omega_{ab}/2 & 0 & \Omega_{qb}/2 & \Omega_{qb}/2 \\0 & \Omega_{qb}/2 & -\omega & 0 \\0 & \Omega_{qb}/2 & 0 & \omega\end{pmatrix}.
\end{align}
Firstly, we note that due to the degeneracy of the $\P$-space, QD-RSPT yields the same results as degenerate RSPT. Interestingly, since the two $\Q$-space states have opposite detunings $\pm \omega$, their contributions to the Stark shift of the ES cancels out. Further, the GS does not couple to the $\Q$-space, and the $\Q$-space states do not couple among themselves. This means that the effective Hamiltonian from adiabatic elimination reads in infinite order
\begin{equation}
    H_\Ae^{(\infty)} = \begin{pmatrix} 0 & \Omega_{ab}/2 \\\Omega_{ab}/2 & 0\end{pmatrix}.
\end{equation}
In other words, all terms beyond first order vanish. In RSPT, we further consider terms beyond the AMP approximation, which are non-zero in this model system. In third order, we have
\begin{equation}
    H_\RS^{(3)} = \begin{pmatrix} 0 & \frac{\Omega_{ab}}{2} \\\frac{\Omega_{ab}}{2} \left( 1 - \frac{ \Omega_{qb}^2}{2\omega^2} \right) & 0\end{pmatrix},
\end{equation}
and in fifth order
\begin{align}
H_\mathrm{RS}^{(5)} = \begin{pmatrix} 0 & \frac{\Omega_{ab}}{2} \\\frac{\Omega_{ab}}{2} \left( 1 - \frac{ \Omega_{qb}^2}{2\omega^2}  - \frac{ \Omega_{qb}^2\Omega_{ab}^2}{8\omega^4}  + \frac{ \Omega_{qb}^4}{4\omega^4} \right) & 0\end{pmatrix}.
\end{align}

As parameters, we choose $\omega = \SI{1}{\au}$, $z_{ab} = \SI{1}{\au}$, and $z_{qb} = \SI{6}{\au}$, which are typical for XUV-driven Rabi oscillations, where the ES couples to a degenerate Rydberg state $\ket{q}$ much more strongly than to the GS. Connecting our model with realistic atoms, $\ket{q}$ represents a manifold of Rydberg states. In real atoms, these are not perfectly degenerate with $\ket{b}$. This complicates the expressions significantly but does not fundamentally alter the physics. 

In Fig.~\ref{fig:model_nonhermitian_H_eff}, we compare the properties and time-evolution due to the $H$ (solid) with those from $H_\Ae^{(\infty)}$ (dashed-dotted), $H_\RS^{(3)}$ (dashed) and $H_\RS^{(5)}$ (dotted). In panel~(a), the dressed energy $E_+$ of the dressed state $\ket{+}$ is compared within the models. Expectedly, $H_\Ae^{(\infty)}$ struggles at high intensities, while $H_\RS^{(n)}$ in third and fifth order both capture the dressing accurately. Note that $E_- = -E_+$ for both $H$ and the effective Hamiltonians.

To gauge the non-Hermiticity of $H_\RS^{(n)}$, we define an effective asymmetry of the absolute value of the off-diagonal elements,
\begin{equation} \label{eq:asymmetrymodel}
    \mathcal{A} \coloneqq \left|\frac{|H_\eff|_{ba}-|H_\eff|_{ab}}{|H_\eff|_{ba}+|H_\eff|_{ab}}\right|,
\end{equation}
and show the results in panel~(b) for $H_\RS^{(3)}$ (dashed) and $H_\RS^{(5)}$ (dotted). Of course, $H_\Ae^{(\infty)}$ (dashed-dotted) is symmetric, so that $\mathcal{A} = 0$. The exact effective asymmetry from $H$ (solid) can be obtained from the exact effective Hamiltonian $H_\eff$, which we obtain using the exact dressed energies $E_\pm$ and the projected dressed states $\ket{\psi_\pm} = P\ket{\Psi_\pm}$ via similarity transform
\begin{equation} \label{eq:exactHeffFloquet}
    H_\eff = \bigl(
        \ket{\psi_-},  \ket{\psi_+}
    \bigr) \begin{pmatrix}
        E_- &  0 \\
        0 & E_+    \end{pmatrix} \bigl(
        \ket{\psi_-},  \ket{\psi_+}
    \bigr)^{-1}.
\end{equation}
The exact $E_\pm$ and $\ket{\Psi_\pm}$ were obtained through diagonalization of $H$.

The time evolution due to $H$ at $E_0 = \SI{0.1}{\au}$ is shown in panel~(c). It is here that the role of the $\Q$-space becomes obvious. It is clearly incorrect to assume no population in the $\Q$-space, which is done in the AMP approximation. The norm in the entire Hilbert space must be conserved; i.e. the full wavefunction $\ket{\Psi(t)} \in \P\oplus\Q$ fulfills $\braket{\Psi(t)|\Psi(t)}=1$. The full wavefunction is made up of its two projections $\ket{\Psi(t)} = P\ket{\Psi(t)}+Q\ket{\Psi(t)}$ and using the reduced wave operator, we may write $\ket{\Psi(t)}=(P+\chi)P\ket{\Psi(t)}$. 

Inspecting panel (c), we observe the GS (yellow) undergoing usual resonant Rabi oscillations. Meanwhile, the coupling of the ES (blue) to $\ket{q}$ (sum of populations of $\ket{q,N-1 \pm 1}$ in green) leads to a transient population in $\Q$. In fact, for this simple model system, with the initial population in the GS, we can identify that the maximum excited state population $\max(|b(t)|^2)$ is determined by $\max(|b(t)|^2) = |(H_\eff)_{ba}/(H_\eff)_{ab}|$. Clearly, $H_\Ae^{(\infty)}$ does not adequately capture the population dynamics of this system. In comparison, already $H_\RS^{(3)}$ provides decent agreement, while $H_\RS^{(5)}$ is excellent in predicting both the Rabi frequency and the magnitude of the excited state.

We note that the full system was evolved with $U(t,0) = \exp(-\i H t)$, which gives rise to a cycle-averaged time evolution, in which the interference due to $\ket{q,N}$ and $\ket{q,N-2}$ is neglected \cite{shirleySolutionSchrodingerEquation1965, chuRecentDevelopmentsSemiclassical1985}. These interferences lead to the appearance of counter-rotating oscillations, where the ES exchanges population with the $\Q$-space on sub-cycle time-scales. Note that this also necessitates accounting for the transient population in the counter-rotating ES $\ket{b, N-1\pm 2}$ \cite{bruhnkeGiantCounterrotatingOscillations2025}. 

Due to the large detuning of $\ket{q, N-1\pm 1}$, their dressing is negligible and all of their population is transient. Transient is however not the same as negligible, which the AMP approximation implies. This population can be predicted by acting on the two-level essential states wavefunction $\ket{\psi(t)} = [a(t), b(t)]^T$ with the reduced wave operator, $Q\ket{\Psi(t)} = \chi \ket{\psi(t)}$. Up to second order in $\chi$, we obtain
\begin{multline}
   \bra{q, N-1\pm 1} \chi \ket{\psi(t)} = - \frac{\Omega_{bv}\Omega_{ba}}{4\omega^2} a(t) \mp \frac{\Omega_{qb}}{2\omega}b(t)  .
\end{multline}
We use this expression to obtain the $\Q$-space population $\sum_\pm | \bra{q, N-1\pm 1} \chi \ket{\psi(t)}|^2$ from the $\P$-space population predicted by $H_\RS^{(3)}$ and $H_\RS^{(5)}$. We find excellent agreement between the exact $\Q$-space population from $H$ and the predictions from RSPT.

\begin{figure*}[t]
    \centering
    \includegraphics{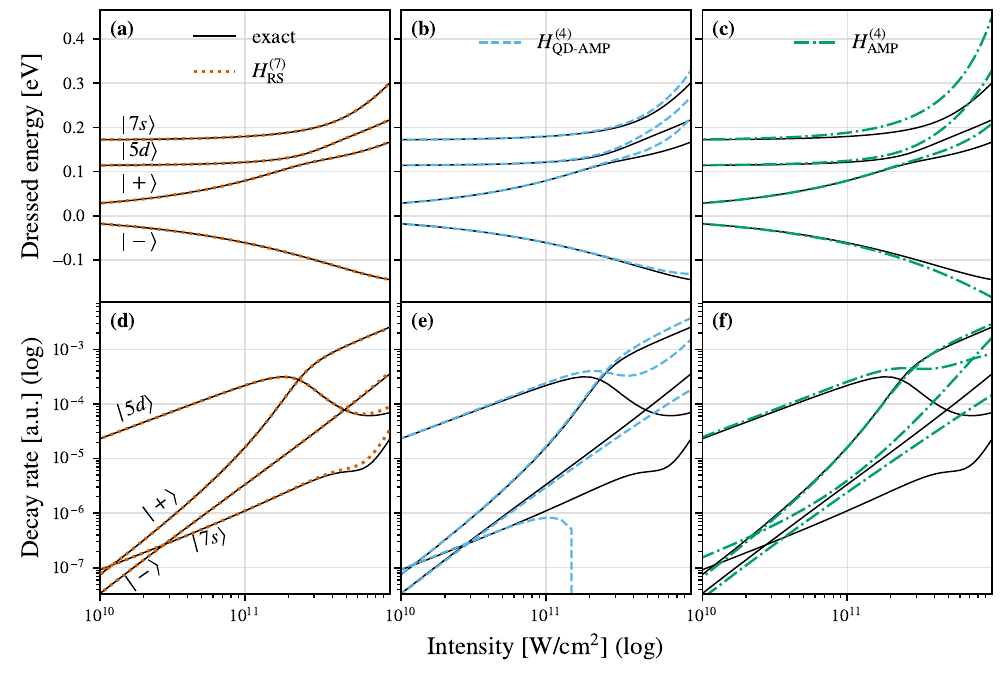}
    \caption{Quasi-energies $E - i\Gamma/2$ of Rubidium in a monochromatic, linearly polarized IR field. The upper row shows the real part of the quasi-energies, $E$; the lower row shows the decay rate $\Gamma$. The black solid lines are exact calculations obtained from Floquet theory. In the left column, the exact quasi-energies are compared to the effective Hamiltonian from QD-RSPT in 7th order (red dotted), see Eq.~\eqref{eq:rspt}. In the middle column, we compare the exact solution to the effective Hamiltonian from QD-RSPT within the pole approximation in 4th order (blue dashed), see Eq.~\eqref{eq:Heff0_nondegenerate}. In the middle column, we compare the exact solution to the effective Hamiltonian from adiabatic elimination in 4th order (green dash-dotted), see Eq.~\eqref{eq:Heff_AE_pert}. We label the dressed states as $\ket{+}$, $\ket{-}$, $\ket{5d}$, and $\ket{7s}$. The labeling of $5d$ and $7s$ refers to their field-free limit; once the $5d$ and $7s$ states are dressed, they mix strongly with other states. }
    \label{fig:rubidium_quasienergies}
\end{figure*}

\subsection{B. Quasi-energies of rubidium in an IR field}

As soon as the essential states are significantly detuned with respect to the optical transition, we should treat the non-degeneracy of the Hamiltonian properly. One system where we can demonstrate the effectiveness of the quasi-degenerate approach is the Rubidium atom, driven by $\SI{800}{\nm}$ linearly polarized light (corresponding to $\SI{1.55}{\eV}$). This showcases the power of the effective Hamiltonian approaches in the more traditional long-wavelength regime of strong-field physics. 

The field couples the $5s$ GS to the $5p$ ES via one photon, creating two dressed states $\ket{\pm}$. The $5p$ state ionizes then via two-photons. Interestingly, the unperturbed $5d$ and $7s$ states are near-resonant to a one-photon transition from $5p$, with detunings between $\SI{0.1}{\eV}$ and $\SI{0.2}{\eV}$. As the intensity increases, the $\ket{+}$ state mixes strongly first with the dressed $5d$ state, and subsequently with the dressed $7s$ state. This constitutes two avoided crossings as the intensity is increased. For this reason, an accurate description of this system requires us to calculate a four-level effective Hamiltonian, with the unperturbed basis $\ket{5s,N}$, $\ket{5p,N-1}$, $\ket{5d,N-2}$, and $\ket{7s,N-2}$.

We construct the full light-matter Hamiltonian $H$ within Floquet theory, where we include all atom-photon states in the basis that are accessible from the essential states via an up to four-photon transition. The required complex-scaled atomic parameters are obtained by solving the TISE for a Rubidium model potential \cite{schweizerMODELPOTENTIALSALKALI1999} with exterior complex-scaling \cite{simonDefinitionMolecularResonance1979}. Thus, we are invoking a single-active electron approximation, in which the 5s valence electron is propagated in an effective potential generated by the inner-shell electrons \cite{schaferThresholdIonizationHigh1993, watsonNonsequentialDoubleIonization1997, mullerBunchingFocusingTunneling1998}. The quasienergies of $H$ are then obtained through sparse matrix diagonalization techniques \cite{virtanenetal.SciPy10Fundamental2020}.

We compare in Fig.~\ref{fig:rubidium_quasienergies} the exact quasienergies $E - \i \Gamma/2$ to those predicted by three different four-level effective Hamiltonians, obtained perturbatively. The upper row shows the real part $E$ while the lower row shows the decay rate $\Gamma$ on a logarithmic scale. In the left column, we present seventh order calculations in QD-RSPT using Eq.~\eqref{eq:rspt}. In the center column, the effective Hamiltonian of the quasi-degenerate AMP approximation, Eq.~\eqref{eq:Heff0_nondegenerate}, is shown in fourth order, while the right column contains results from the traditional AMP effective Hamiltonian, Eq.~\eqref{eq:Heff_AE_pert}, also in fourth order. For the latter, we choose the free parameter $E_0$ to be the center of the unperturbed $\P$-space spectrum according to common practice. We choose fourth order for $H_{\QDAE}$ and $H_\Ae$ since higher orders are not common in the literature \cite{beersExactSolutionRealistic1975, holtTimeDependencesTwo1983, dorrTimeEvolutionTwophoton1997, zhangEffectNonresonantStates2022, olofssonPhotoelectronSignatureDressedatom2023}. Further, higher orders do not substantially improve the agreement. %, except for the decay rate of the dressed 7s state of $H_\QDAE$ in panel (e). 
In contrast, seventh-order in QD-RSPT is required to get good agreement of the decay rates with the exact solution. 

QD-RSPT yields accurate results for all calculated intensities. Notably, even the complex ionization dynamics at high intensities are reproduced rather well. Applying our quasi-degenerate AMP approximation yields very good results both for the quasi-energies and decay rates at low intensities. Beyond the first avoided crossing of $\ket{+}$ with the dressed $\ket{5d}$, the accuracy quickly worsens, even predicting unphysical positive decay rates for one of the dressed states (this can be improved by including higher-order terms). The AMP approximation is unsuited for this problem, most notably for predicting the ionization dynamics, the dressed energies are however predicted reasonably well up to the first avoided crossing. 

At $I_0 < \SI{e11}{\watt\per\square\cm}$ we can accurately model the two-level dynamics of the two dressed states $\ket{\pm}$. For $\SI{e11}{\watt\per\square\cm} < I_0 < \SI{e12}{\watt\per\square\cm}$, the two avoided crossings lead to an effective four level system. Beyond $I_0 > \SI{e12}{\watt\per\square\cm}$, the coupling between $\P$ and $\Q$ becomes so substantial that the effective Hamiltonian approach breaks down, and many-level dynamics will dominate, marking the onset of strong-field effects.

\subsection{C. Quasi-energies of helium in an XUV field}

%With the invention of seeded free-electron lasers, coherent non-linear processes in the high-frequency regime have recently become accessible to study, such as Rabi oscillations in helium \cite{nandietal.ObservationRabiDynamics2022, nandiGenerationEntanglementUsing2024, richteretal.StrongfieldQuantumControl2024}. 
In the XUV and X-ray regime, the detuning of intermediate states is usually very large, thus allowing for large intensities upwards of $\SI{e14}{\watt\per\square\cm}$. At these large intensities, non-linear effects, such as non-resonant two-photon ionization from the ground state, or counter-rotating transitions, strongly shape the dynamics. Here, QD-RSPT provides clear instructions in how to treat these effects systematically, order by order. 

\begin{figure}[t]
    \centering
    \includegraphics{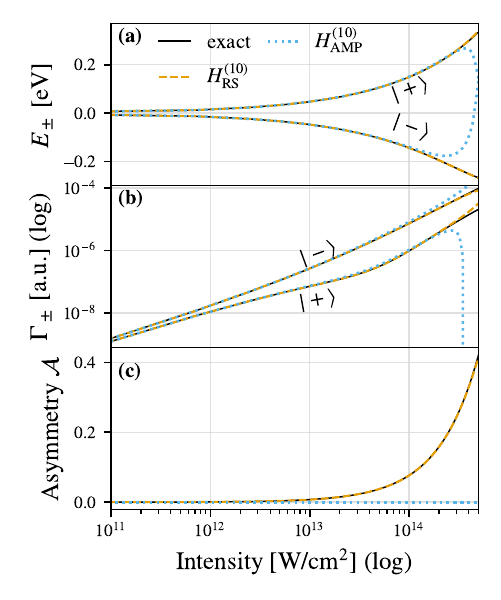}
    \caption{Quasi-energies $E_\pm - \i\Gamma_\pm/2$ of Helium in a monochromatic, linearly polarized XUV field, resonant with the $1s^2\leftrightarrow 1s3p$ transition. Panel (a) shows the real part of the quasi-energies, $E_\pm$, and panel (b) the decay rate $\Gamma_\pm$. The black solid lines are exact calculations obtained from Floquet theory. We compare the exact quasi-energies to the effective Hamiltonian from RSPT in 10th order (yellow dashed), see Eq.~\eqref{eq:rspt}, and to the AMP effective Hamiltonian in 10th order (blue dotted), see Eq.~\eqref{eq:Heff_AE_pert} (since the field is exactly degenerate, $H_\Ae = H_{\QDAE}$). We label the dressed states as $\ket{+}$ and $\ket{-}$. In panel (c), we show the asymmetry of the Hamiltonian, where the asymmetry is defined in Eq.~\eqref{eq:asymmetrymodel}. The exact asymmetry was obtained by calculating the effective Hamiltonian from diagonalization of the Floquet Hamiltonian, see Eq.~\eqref{eq:exactHeffFloquet}. }
    \label{fig:helium_quasienergies}
\end{figure}

As an example system, we study the (1+1)-REMPI process in helium for the resonant $1s^2\leftrightarrow 1s3p$ transition in Fig.~\ref{fig:helium_quasienergies}. Therefore, QD-RSPT is equivalent to degenerate RSPT. We benchmark the RSPT calculations against exact solutions, obtained via Floquet theory. The complex-scaled atomic parameters (energies and dipole moments) are obtained by diagonalizing the Hamiltonian from the configuration interaction singles method \cite{foresmanSystematicMolecularOrbital1992, dreuwSingleReferenceInitioMethods2005}. We construct the full light-matter Floquet Hamiltonian $H$ within Floquet theory, where we include all atom-photon states in the basis that are accessible from the essential states via an up to five-photon transition. Note that we do not account for doubly excited states. Although we expect this to be a good approximation for these parameters, it may be a worthwhile future endeavor to study transitions to doubly excited states. The effective Hamiltonian of RSPT is then obtained in tenth order. We obtain a $2\times 2$ complex non-symmetric effective Hamiltonian of the form
\begin{align}
    H_\RS = \begin{pmatrix}
        S_a - \frac{\i}{2}\gamma_a & (\Omega_{ab} + \i \beta_{ab}) / 2 \\
        (\Omega_{ba} + \i \beta_{ba})/2 & \delta + S_b - \frac{\i}{2} \gamma_b
    \end{pmatrix}, \label{eq:cnsHeff}
\end{align}
where $S_{a/b}$ are the Stark shifts, $\delta$ the detuning, $\gamma_{a/b}$ the ionization rates, and $\Omega_{ab/ba}$ and $\beta_{ab/ba}$ the real and imaginary parts of the effective Rabi frequency. Diagonalization of $H_\eff$ yields the quasienergies $E_\pm - \i \Gamma_\pm /2$ of the two dressed states $\ket{\pm}$.

In Fig.~\ref{fig:helium_quasienergies}, we show $E_\pm$ and $\Gamma_\pm$ in panel (a) and (b) respectively. We see that at intensities beyond $\SI{2e14}{\watt\per\square\cm}$, the tenth order AMP effective Hamiltonian, $H_\Ae$, ceases to match the exact results, while $H_\RS$ agrees excellently with the Floquet calculations. In panel (c), we plot the effective asymmetry of the effective Hamiltonian, defined by Eq.~\eqref{eq:asymmetrymodel}. We further obtain the exact effective Hamiltonian non-perturbatively from Floquet theory through sparse matrix diagonalization techniques. By applying Eq.~\eqref{eq:exactHeffFloquet}, we obtain the exact effective Hamiltonian via similarity transform. 
%. By calculating the exact eigenenergies $E_\pm$ and the corresponding full eigenstates $\ket{\Psi_\pm} \in \P\oplus\Q$, we can project the eigenstates onto $\P$ with $\ket{\psi_\pm} \coloneqq P\ket{\Psi_\pm}$ and then obtain the exact effective Hamiltonian via similarity transform:
%
%\begin{equation} \label{eq:exactHeffFloquet}
%    H_\eff = \bigl(
%        \ket{\psi_-},  \ket{\psi_+}
%    \bigr) \begin{pmatrix}
%        E_- &  0 \\
%        0 & E_+    \end{pmatrix} \bigl(
%        \ket{\psi_-},  \ket{\psi_+}
%    \bigr)^{-1}.
%\end{equation}
%
It is clear from panel (c) that already beyond $\SI{e13}{\watt\per\square\cm}$, the non-orthogonality of the $\P$-space becomes non-negligible, and $H_\RS$ needs to be employed to account for it, even if the eigenenergies of $H_\Ae$ stay accurate for another order of magnitude of the intensity. The source and physical implication of the effective asymmetry was discussed earlier this section, see especially Fig.~\ref{fig:model_nonhermitian_H_eff}(c). In our helium $1s^2 \leftrightarrow 1s3p$ system at high intensities, the $1s3p$ state couples strongly to nearby $s$- and $d$-Rydberg states in $\Q$, while the ground state couples only negligibly to $\Q$. This asymmetric coupling is encoded in the effective asymmetry Eq.~\eqref{eq:asymmetrymodel}, and is the origin for non-reciprocal counter-rotating oscillations, see Ref.~\cite{bruhnkeGiantCounterrotatingOscillations2025}. Meanwhile, $H_\Ae$ is complex-symmetric by design; therefore its asymmetry is 0 for all intensities.

\section{V. Conclusion}

Coherent processes in light-matter interaction are routinely described via essential states approaches, where adiabatic elimination, the Markov approximation, and the pole approximation are the standard approaches used to obtain an effective Hamiltonian $H_\Ae$ acting in a Hilbert subspace $\P$. However, at high intensities, $H_\Ae$ fails to describe the physics, both quantitatively and qualitatively. One weak point is that traditional methods such as adiabatic elimination, the pole approximation, and the Markov approximation, assume implicitly that the incoming field is nearly resonant, i.e. that the unperturbed states in $\P$ are degenerate. This makes the methods sensitive to an arbitrary energy shift. Most importantly, the adiabatic elimination of nonessential states forces all states in $\P$ to couple with the same strength to the nonessential states in $\Q$. We have demonstrated that this assumption is untenable. Against this backdrop, QD-RSPT emerges as a versatile and accurate tool that properly accounts for the detunings of essential states, and incorporates the asymmetric coupling from $\P$ to $\Q$ by design. Mathematically, this is encapsulated by the non-orthogonal eigenvectors of the effective Hamiltonian $H_\RS$. Through QD-RSPT, we have shown that we can describe intricate quasi-energy structures in intense fields, both in the traditional low-frequency regime, and in the high-frequency domain that was recently made accessible through seeded free-electron lasers. %These include the description of avoided crossings with complicated decay dynamics in rubidium, and the strong asymmetry in the coupling from $\P$ to $\Q$ in strongly coupled helium, which have shown to lead to the emergence of giant counter-rotating oscillations \cite{bruhnkeGiantCounterrotatingOscillations2025}. Exploring the implications of these features for time-dependent observables will be the topic of upcoming work. \textcolor{red}{Better now?}

%Through QD-RSPT, we can describe intricate coherent phenomena in intense fields 

%We showed that QD-RSPT yields quantitative agreement with exact calculations in model systems and real atoms.

Even at moderate intensities, where the corrections beyond the AMP approximation are not required, it can still be crucial to incorporate the non-degeneracy in $\P$. To this end, we have introduced the quasi-degenerate AMP approximation, in which all essential states are treated on an equal footing. We underpinned this through simulations in Rubidium with an IR field, in which the quasi-degenerate AMP approximation clearly outperformed the traditional AMP approximation, at no added complexity in the perturbative series.

For higher intensities, the coupling between $\P$ and $\Q$ becomes so strong that QD-RSPT becomes indispensable. Such coupling was explored in a model system, where an excited state coupled to a degenerate Rydberg state. This model connects QD-RSPT to the recent predictions of giant non-reciprocal counter-rotating oscillations \cite{bruhnkeGiantCounterrotatingOscillations2025}. 
The predictive power of QD-RSPT comes at a price: Already at low orders ($n\geq 4$), the number of terms in the perturbative expansion becomes so large that purely analytical endeavors with pen and paper are unproductive. Thus, QD-RSPT does not make small scale essential states approaches obsolete. Instead, QD-RSPT is useful especially in an intermediate regime, where the physical processes of interest are still due to only a handful of states, but the intensity is so high that more sophisticated approaches than the (quasi-degenerate) AMP approximation are required.

%In Section~II, we summarized the textbook approach to QD-RSPT, as proposed by Lindgren \cite{lindgrenRayleighSchrodingerPerturbationLinkeddiagram1974, lindgrenAtomicManyBodyTheory1982}, showcasing in particular the emergence of the asymmetric coupling from $\P$ to $\Q$, and the relation to the degenerate perturbation theory. We discussed the elimination of the continuum through complex-scaling techniques and showcased Floquet theory as a method to obtain a time-independent light-matter interaction Hamiltonian. We compared QD-RSPT in Section~III with the traditional adiabatic elimination, leading us to define a quasi-degenerate adiabatic elimination procedure that treats all essential states on an equal footing, no matter the detuning of the incoming field. We further extended the usual validity condition of adiabatic elimination for a single nonessential state to be valid for arbitrary $\Q$-spaces.

It is important to note that the calculation of $H_\RS$ from the full Hamiltonian is computationally very cheap. %The same applies to the numerical propagation in a few-level subspace, even with the time-dependent effective Hamiltonians used in Section~IV and Section~V. 
Thus, QD-RSPT allows us to scan large parameter spaces of frequency and field strength. To provide an outlook, this enables the efficient calculation of experimentally relevant observables such as photoelectron spectra for systems that would otherwise require the costly propagation of the full TDSE, or diagonalization of large matrices in Floquet theory \cite{dorrMultiphotonProcessesIntense1990}. Other observables that can and should be modeled include the absorption and stimulated emission, for which we expect that the theory of effective operators \cite{suzukiEffectiveOperatorsTimeIndependent1995} will prove helpful. More recently, the field of intense quantum optics was established \cite{gorlachQuantumopticalNatureHigh2020}. In this field, new theoretical tools are being developed, in which the single-atom response to quantum light is modeled through appropriate averaging over semiclassical calculations \cite{gorlachHighharmonicGenerationDriven2023, gonzalez-mongeHighharmonicGenerationDriven2025, langeHierarchyApproximationsDescribing2025}. If resonant atomic transitions are targeted with intense quantum light, we expect QD-RSPT to be of great use in this area to make quantitative studies numerically feasible.

\section{Acknowledgements}

We acknowledge Edvin Olofsson for insightful discussions. JMD acknowledges support from the Knut and Alice Wallenberg Foundation: 2024.0212 and the Swedish Research Council: 2024-04247.

\appendix

\section{Appendix: Effective Hamiltonians from adiabatic elimination, the Markov approximation, and the pole approximation}

\subsection{A. Energy domain: The pole approximation}

The pole approximation can be derived in several ways. We choose a conceptually simple path, starting with the TISE $H\ket{\Psi} = E\ket{\Psi}$. We obtain a $\P$-space and $\Q$-space TDSE by inserting $P+Q = 1$ before $\ket{\Psi}$, and then projecting from the left with either $P$ or $Q$:
\begin{align}
    PHP \ket{\Psi} + PHQ \ket{\Psi} &= E P \ket{\Psi}, \label{eq:PprojTISE}\\
    QHQ \ket{\Psi} + QHP \ket{\Psi} &= E Q \ket{\Psi}. \label{eq:QprojTISE}
\end{align}
Due to the properties of the projectors $P$ and $Q$, we have $PHQ = PVQ$ and $QHP = QVP$. The $\P$-space TDSE, Eq.~\eqref{eq:PprojTISE}, still depends on $Q\ket{\Psi}$. By rearranging Eq.~\eqref{eq:QprojTISE}, we can express $Q\ket{\Psi}$ as a function of $P\ket{\Psi}$, which we insert into Eq.~\eqref{eq:PprojTISE} in order to obtain a TISE for the $\P$-space dynamics, 
\begin{align}
    \left( PHP + PV \frac{Q}{E - QHQ} VP \right) \ket{\Psi} = E P \ket{\Psi}. \label{eq:energydep_TISE}
\end{align}
Here, we identify the Bloch-Horowitz Hamiltonian \cite{blochDeterminationPremiersEtats1958} (also called the Feshbach operator \cite{feshbachUnifiedTheoryNuclear1962})
\begin{equation}
    H_\BH(E) = PHP + PV \frac{Q}{E - QHQ} VP,
\end{equation}
which is an energy-dependent effective Hamiltonian that appears in Brillouin-Wigner perturbation theory \cite{lowdinNoteQuantumMechanicalPerturbation1951, lindgrenAtomicManyBodyTheory1982}. Since $H_\BH(E)$ is energy-dependent, Eq.~\eqref{eq:energydep_TISE} is a non-linear eigenvalue problem and must be solved self-consistently one eigenenergy at a time \cite{lindgrenAtomicManyBodyTheory1982}. Note Eq.~\eqref{eq:energydep_TISE} is formally equivalent to the full TISE, given that no approximations have been made thus far. 

The pole approximation is arguably the most simple method to eliminate the energy dependence of $H_\BH(E)$, and has been used to great effect, for example to explain phenomena in resonant multiphoton ionization \cite{beersExactSolutionRealistic1975, hansonTheoryCoreResonantIonization1995, olofssonPhotoelectronSignatureDressedatom2023}, or to predict the bound-states dynamics in a low-energy subspace \cite{vothAdiabaticallyReducedCoupled1986}. It can be regarded as the implementation of adiabatic elimination in the energy domain \cite{brionAdiabaticEliminationLambda2007}. The energy-dependence of $H_\BH(E)$ is weak if the $\P$-space eigenvalues are close to another and well-separated from the $\Q$-space spectrum. In this case, the pole approximation consists of evaluating $H_\BH(E)$ at a fixed energy $E_0$ that lies approximately in the centre of the $\P$-space spectrum \cite{cohen-tannoudjiAtomPhotonInteractionsBasic1998}:
\begin{equation} \label{eq:Heff0}
    H_\Ae = H_\BH(E_0) = PHP + PV \frac{Q}{E_0 - QHQ} VP. 
\end{equation}
The optimal choice of $E_0$---where the eigenvalues of $H_\Ae$ are closest to the true eigenvalues---is impossible to determine \textit{a priori}.

\subsection{B. Time domain: Adiabatic elimination and the Markov approximation}

In analogy with the time-independent case, we partition the TDSE for a time-independent Hamiltonian, $\i \ddt \ket{\Psi(t)} = H\ket{\Psi(t)}$, into
\begin{align}
\i \ddt P\ket{\Psi(t)} &= PH P \ket{\Psi(t)} + PV Q \ket{\Psi(t)}, \label{eq:equationforsubspacep_paulisch}\\
\i \ddt Q\ket{\Psi(t)} &= QV P \ket{\Psi(t)} + QH Q \ket{\Psi(t)}. \label{eq:equationforsubspaceq_paulisch}
\end{align}
In the Markov approximation, we first formally solve Eq.~\eqref{eq:equationforsubspaceq_paulisch}, 
\begin{equation} \label{eq:qpsi_paulisch}
Q\ket{\Psi(t)} = -\i \int_0^t \d t' \, \e^{-\i QH Q (t-t')} QV P\ket{\Psi(t')}
\end{equation}
in order to obtain the $\P$-space TDSE
\begin{multline} \label{eq:TDSE_pspace}
\i \ddt P\ket{\Psi(t)} =  PH P \ket{\Psi(t)} \\ - \i  PV Q\int_0^t \d t' \, \e^{-\i QH Q (t-t')} QV P\ket{\Psi(t')}.
\end{multline}
We now apply the Markov approximation $P\ket{\Psi(t')} \approx P\ket{\Psi(t)}$. This amounts to neglecting the ``history'' of $P\ket{\Psi(t')}$ before $t$. The integral becomes solvable, yielding
\begin{equation} \label{eq:qpsi_integral_solved}
\int_0^t \d t' \, \e^{-\i QH Q (t-t')} = \frac{1 - \e^{-\i QH Q t}}{\i QHQ} \approx -\frac{\i}{QHQ}
\end{equation}
where we coarse-grained $\e^{-\i QHQ t} \approx 0$. The physical reasoning here is that the evolution due to the $\Q$-space states is highly non-resonant, therefore only giving rise to rapid oscillations which average out. Coarse-graining is of course essential to arrive at a true effective Hamiltonian, which drives the dynamics only due to its own eigenvalues, the dressed energies. 

Inserting in Eq.~\eqref{eq:equationforsubspacep_paulisch} yields the usual effective Hamiltonian from adiabatic elimination \cite{paulischAdiabaticEliminationHierarchy2014},
\begin{equation} \label{eq:Heff0paulisch}
H_\Ae = PH P - PV\frac{Q}{QHQ}VP.
\end{equation}
Note that $H_\Ae$ is also obtained swiftly when setting the driving force zero, $\i \ddt Q\ket{\Psi(t)} \approx 0$. This canonical approach to adiabatic elimination amounts to only considering the DC response \cite{allenBroadeningSaturationNphoton1982}. Of course, if we shift the original Hamiltonian by $H \rar H - E_0$, this will not affect the dynamics of the full system, since it corresponds to a unitary transformation. In adiabatic elimination however, it will shift the $\Q$-space resolvent, leading to
\begin{equation} \label{eq:Heff0energyoffset}
    H_\Ae = P(H-E_0)P + PV\frac{Q}{E_0 - QHQ} VP.
\end{equation}
which generates a different dynamics than Eq.~\eqref{eq:Heff0paulisch}. Of course, the energy shift $PHP \rar P(H-E_0)P$ is still inconsequential, so that Eq.~\eqref{eq:Heff0} and~\eqref{eq:Heff0energyoffset} give rise to the same physics.

We note finally that Paulisch \textit{et al.} have proposed a higher-order Markov approximation leading to an effective Hamiltonian beyond the AMP approximation. Their higher-order effective Hamiltonian is notably non-perturbative, making it less suitable for the description of large atomic systems with thousands of basis states. We have previously shown how the higher-order Markov approximation is equivalent to a higher-order pole approximation in first order \cite{bruhnkeGiantCounterrotatingOscillations2025}. Connecting it to QD-RSPT, the effective Hamiltonian from the higher-order Markov approximation is a partial resummation of specific terms of the Rayleigh-Schrödinger perturbative expansion. By constructing an inductive proof, we have managed to show that the infinite-order Markov approximation is equivalent to the Krenciglowa-Kuo iterative procedure from nuclear structure theory \cite{krenciglowaConvergenceEffectiveHamiltonian1974, takayanagiEffectiveHamiltonianExtended2011}; it is however outside the scope of this work. If the procedure converges, the exact effective Hamiltonian $H_\eff$ is obtained.

\bibliography{MyLibrary}

\end{document}